# The dominancy of damping like torque for the current induced magnetization switching in Pt/Co/W multilayers

Zelalem Abebe Bekele, Kangkang Meng*, Yong Wu, Jun Miao, Xiaoguang Xu, Yong Jiang*

*School of Materials Science and Engineering, University of Science and Technology Beijing,*

*Beijing 100083, China*

**Abstract**

Two classes of spin-orbit coupling (SOC) mechanisms have been considered as candidate sources for the spin orbit torque (SOT): the spin Hall Effect (SHE) in heavy metals with strong SOC and the Rashba effect arising from broken inversion symmetry at material surfaces and interfaces. In this work, we have investigated the SOT in perpendicularly magnetized Pt/Co/W films, which is compared with the results in Pt/Co/AlO$_x$ films. Theoretically, in the case of the asymmetric structure of trilayers with opposite sign of spin Hall angle, both damping like torque and field like torque due to the SHE and the Rashba effect will be enhanced. Using the harmonic measurements, we have characterized the effective fields corresponding to the damping like torque and the field like torque, but we have found the dominancy of damping like torque in the Pt/Co/W films. It is much different from the results in the Pt/Co/AlO$_x$ films, in which both the damping like torque and the field like torque are strong.

*Authors to whom correspondence should be addressed:

*kkmeng@ustb.edu.cn

*yjiang@ustb.edu.cn

## 1. Introduction

Magnetization reversal by an electric current is crucial issue for spintronic memory and logic devices, such as spin transfer torque magnetic random access memory, in which magnetization switching are occurred in a spin valve structure [1]. Recently, another type of torques namely spin orbit torque (SOT) has attracted great interest as they can control the magnetization in heavy metal (HM)/ferromagnetic (FM) heterostructures with perpendicular magnetic anisotropy (PMA) [2-13]. Typically, in HM/FM/oxide trilayers with PMA, the charge current passing through the HM can produce pure spin current due to spin Hall effect (SHE), which exerts a torque on the magnetization. Hence, the sign and magnitude of SOT are essentially determined by the spin Hall angle in the HM. On the other hand, the SOT can also arise from the interfacial Rashba effect, for which the accumulated spins at the interface can force the moments to change its direction by direct exchange coupling. These two mechanisms are all related to strong spin orbit coupling (SOC) [14-17]. Recent theoretical studies showed that both the SHE and the Rashba effect can generate damping like torque and field like torque. Correspondingly, the two kinds of torques yield to the damping like effective field $H_D \sim \sigma \times m$ and the field like effective field $H_F \sim \sigma$, respectively, where m and σ are unit vectors of the magnetization in the FM layer and non-equilibrium spin polarization direction in the HMs. Particularly, the direction of σ depends on the sign of spin Hall angle $\theta_{SH}$ of HMs.

Currently, most studies have focused on SOT in ultrathin FMs sandwiched between two HM layers, such as Pt/CoNiCo/Pt, Pt/Co/Ta, Pt/Co-Ni/W, and Pt/Co/Pt [18-23], in which both of the HM layers have contributions to the spin accumulation at the interfaces. Because the efficiency of SOT relates directly to the magnitude of effective spin Hall angle ($\theta_{SH}$), it is considered that the overall $\theta_{SH}$ will be enhanced if the sign of $\theta_{SH}$ in the top and bottom HM layer is opposite. In this work, we studied the SOT in Pt/Co/W trilayers, in which the sign of $\theta_{SH}$ is positive in Pt while the value is negative in W. Theoretically, in the case of asymmetric structures with opposite sign of spin Hall angle, both the SHE and the Rashba effect will be enhanced. Correspondingly, the two kinds of torques will be increased. However, we found large damping like torque in the Pt/Co/W multilayers as compared with the field like torque. The results were much different from the Pt/Co/AlO$_x$ films, in which both the damping like torque and the field like torque are comparable.

## 2. Results and discussion

W(1.2)/Pt(6)/Co(1)/AlO$_x$(3)/Pt(1) and W(1.2)/Pt(6)/Co(1)/W(2) (in nanometer from bottom to top) were prepared by magnetron sputtering system on Si/SiO$_2$ substrate as shown in Figure 1(a). The two samples are referred to as sample I and sample II. After the deposition process, we annealed both of the samples at 250 ℃ for 40 min at the base pressure of $4.2\times10^{-5}$ Pa to enhance the PMA. The thin films were patterned into to 15 μm×80 μm Hall bar structures by electron beam lithography (EBL). Figure 1(b) shows the schematic of the Hall bar along with the definition of the coordinate system. The magnetic hysteresis loops (M-H curves) as shown in Figure 1(c) indicate strong PMA in both of the two samples. On the other hand, the Hall resistances (R$_H$) as a function of the out-of-plane magnetic field (H) were measured at a source current of 0.1 mA as shown in Figure 1(d). They also present strong PMA of the two samples.

We measured the SOT induced magnetization switching by applying a pulsed current with the width of 50 μs, and the resistance was measured after a 16 μs delay under an external magnetic field along the x-direction. Figure 2 (a) and (b) showed the current induced magnetization switching with varying applied in-plane field for sample I and II, respectively. Reductions of the critical switching currents were observed as the external field increased for both of the two samples. Furthermore, with applying appropriate field (~1000 Oe) the magnetizations are fully switched. It can be proved by the changes of the Hall resistance of the two samples, which are consistent with the results in $R_H$-H measurements as shown in Figure 1(d). Considering the resistivity of the W layer is much larger than that of the Pt layer and the $AlO_x$ layer should be insulating, here we roughly took the thickness of Co and Pt layers into account when determining the switching current density ($J_c$) of the two samples. Therefore, with applying the magnetic field of 3000 Oe (fully magnetization switching), the $J_c$ in the Pt/Co/$AlO_x$ and Pt/Co/W films are $(3.8\pm0.051)\times10^{11}$ $Am^{-2}$ and $(2.04\pm0.0044)\times10^{11}$ $Am^{-2}$, respectively.

The harmonic measurements were determined by applying the sinusoidal AC current, for sample I the frequency is 365 Hz and the amplitude is ranged from 2.0 to 8.0 mA, for sample II the frequency is 315 Hz and the amplitude is ranged from 2.0 to 2.6 mA. The first ($V_\omega$) and second ($V_{2\omega}$) harmonic Hall voltages are detected using two lock-in amplifier systems at the same time by sweeping the longitudinal ($H_x$) and transverse ($H_y$) fields. Before the harmonic measurements, we have applied a large out-of-plane external field to the two samples, which remain saturated after the field is turned off. Here, we will take the example of the harmonic measurements for sample I with 5.0 mA and for sample II with 2.0 mA respectively as shown in Figure 3, in which the results are measured with out-of-plane magnetization component $M_Z>0$. Then, the values of the damping like field ($H_D$) and field like effective field ($H_F$) of the two samples can be evaluated using the following relation [24]:

$$H_{D(F)} = -2\frac{H_{L(T)} \pm 2\xi H_{T(L)}}{1-4\xi^2} \quad (1)$$

Where $\xi$ defined as the ratio of planar Hall effect (PHE) resistance and anomalous Hall effect (AHE) resistance and $\pm$ sign is the direction of magnetization pointing ( $\pm M_z$) along out-of-palne axis [13, 24]. The longitudinal ($H_L$) and transverse effective field ($H_T$) can be calculated using the following relation:

$$H_{L(T)} = \frac{\partial V_{2\omega}/\partial H_{X(Y)}}{\partial^2 V_{\omega}/\partial H_{X(Y)}^2} \quad (2)$$

To measure the PHE resistance $R_{PHE}$, a 30000 Oe strong in-plane magnetic field is applied to saturate the magnetization of the samples [9, 10, 13, 24]. Then, the $R_H$ was measured by rotating the magnetic field in the plane of sample to rotate the magnetization along the $\alpha$ direction. The planar Hall measurement data can be fitted using the equation $R_H = R_{PHE}\sin(2\alpha+\varphi_1) + R_o\sin(\alpha+\varphi_2) + C$ to obtain the value of $R_{PHE}$ [9, 10, 13], $\varphi_1$ and $\varphi_2$ are the offset angles in our measurement setup, C is the offset value in $R_H$ and the second term ($R_o\sin\alpha$) is the contribution of anomalous Hall effect due to small misalignment of external magnetic field to the sample plane. The experimental data and fitted lines for $R_H$-$\alpha$ of the two samples are shown in Figure 4(a) and (b) respectively. Then the values of $\xi$ in the two samples can be determined to be 1.02 and 1.405 for sample I and II, respectively. According to the Equation (1), the $H_D$ and $H_F$ of the two samples can be finally determined and the results are shown in Figure 4(c) and (d). The effective fields vary linearly with the current amplitude (I), indicating that the effects of Joule heating are negligible in the measured current range. However, the values of $H_F$ in the Pt/Co/W films are nearly ~88% smaller than $H_D$, indicating that the current induced magnetization switching of this sample has mostly depended on the damping like torque. Finally, the spin Hall angle $\theta_{SH}$ were calculated using $\left(\theta_{SH} = \frac{2(H_D|e|M_s t_{FM})}{\hbar|J_c|}\right)$, where e is charge of an electron, $M_s$ is saturation magnetization, $t_{FM}$ is thickness of ferromagnetic layer, $J_c$

is applied current density, and $\hbar$ is reduced Planck constant [24]. The effective $\theta_{SH}$ up to 0.12 and 0.3004 were determined for sample I and sample II, respectively. Therefore, the effective $\theta_{SH}$ will be enhanced when the Co layer is sandwiched between two HM layers Pt and W with opposite signs of $\theta_{SH}$, and the damping like torque play a dominant role.

In the previous studies on SOT of HM/FM/oxide trilayers, the spin density generated by the inverse spin galvanic effect exerts a torque on the magnetization, which is attributed to interfacial Rashba SOC [5-9]. A major difficulty to identify the physical origin of the SOT is that the SHE also plays an important role in magnetic multilayers. Recently, Haney *et al*. have developed semi-classical models for electron and spin transport in bilayer nanowires with a ferromagnetic layer and a nonmagnetic layer with strong SOC [25]. They have proved that the damping like torque is typically derived from the models describing the bulk SHE and the spin transfer torque, and the field like torque is typically derived from a Rashba model describing interfacial SOC. For the Rashba effect, an internal electric field gradient will arise at the interfaces along the direction of symmetry breaking. The direction of Rashba effect induced spin accumulation as well as an effective magnetic field will be along ***E*×*P***, where ***E*** is the internal electric field, ***P*** is the electron momentum [26]. The direction of the ***E*** for a given interface can be obtained by considering the differences in the work function (Φ) of the two materials at the interface. The work functions of Pt, Co, W and AlO$_x$ are ~5.65 eV, 5.0 eV, 4.55 eV and 3.2 eV [27]. The smaller difference of work function in Pt/Co/W as compared with that in Pt/Co/AlO$_x$ films will decrease the spin accumulation from the Rashba effect. On the other hand, to obtain the β-phase of W layer a dc sputtering power of 8 W was applied with the growth rate of 0.03nm/s, and the cubic lattice constant of β-W is 0.5046 nm [28]. For the hexagonal lattice of Co, the in-plane lattice constant is about 0.2507nm, which is almost half of that in β-W. Therefore, we propose that the asymmetrical properties in Pt/Co/W are not as obvious as that in Pt/Co/AlO$_x$ films and the Rashba effect becomes smaller. Finally, as the resistivity of W is much smaller than that of AlO$_x$, the shunting effect is smaller in Pt/Co/W films. Thus, for a given current density, spin accumulation from the bulk SHE will be larger in Pt/Co/W films, for which the damping like torque due to SHE will play a dominant role.

## 3. Conclusion

In conclusion, we have investigated the SOT in perpendicularly magnetized Pt/Co/AlO$_x$ and Pt/Co/W films. Using the harmonic measurements, we have characterized the effective fields corresponding to the damping like torque and the field like torque and we have found obvious damping like torque in the Pt/Co/W films while the field like torque was small. It is much different from the results in the Pt/Co/AlO$_x$ films, in which both the damping like torque and the field like torque are strong. According to the weak asymmetrical properties in Pt/Co/W, we propose the Rashba effect in this film is weak.

**ACKNOWLEDGEMENT**

This work was partially supported by the National Basic Research Program of China (2015CB921502), the National Science Foundation of China (Grant Nos. 51731003, 61404125, 51471029, 51671019, 11574027, 51501007, 51602022, 61674013, 51602025), and the Fundamental Research Funds for the Central Universities (FRF-GF-17-B6).

**Figure captions**

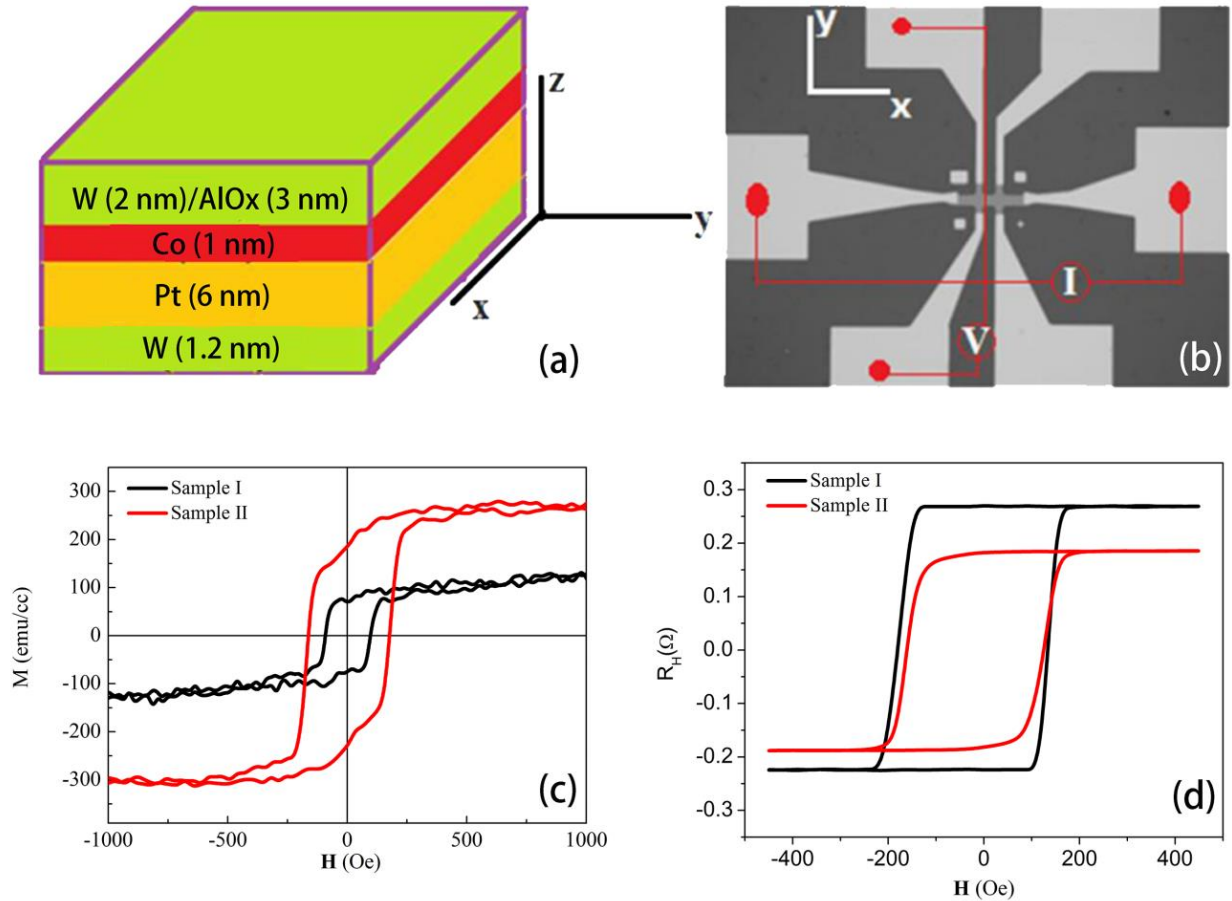

Figure 1 (a) Schematic structure of the multilayers. (b) SEM image of a Hall bar and measurement setup. (c) The M-H curves of the two samples. (d) $R_H$-H curves of the two samples.

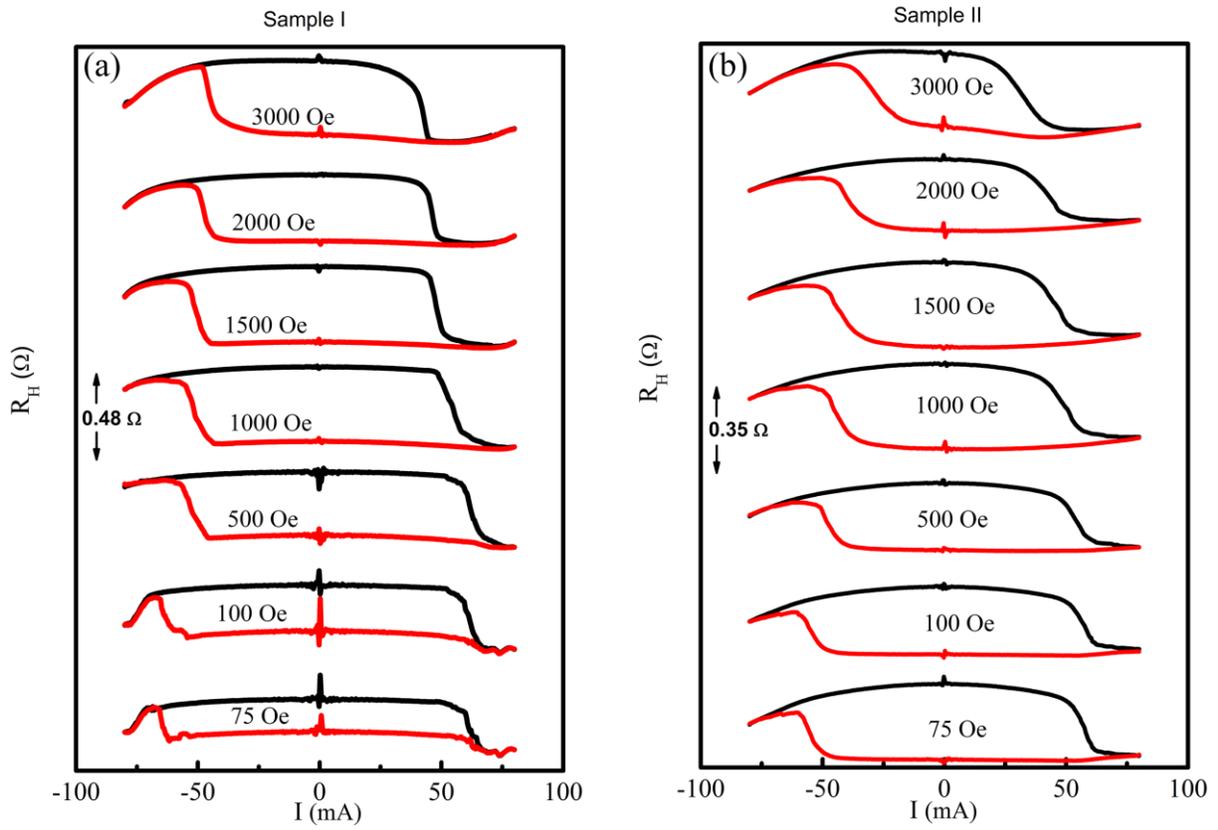

Figure 2 $R_H$-I curves in sample I (a) and sample II (b) with varying in-plane magnetic field.

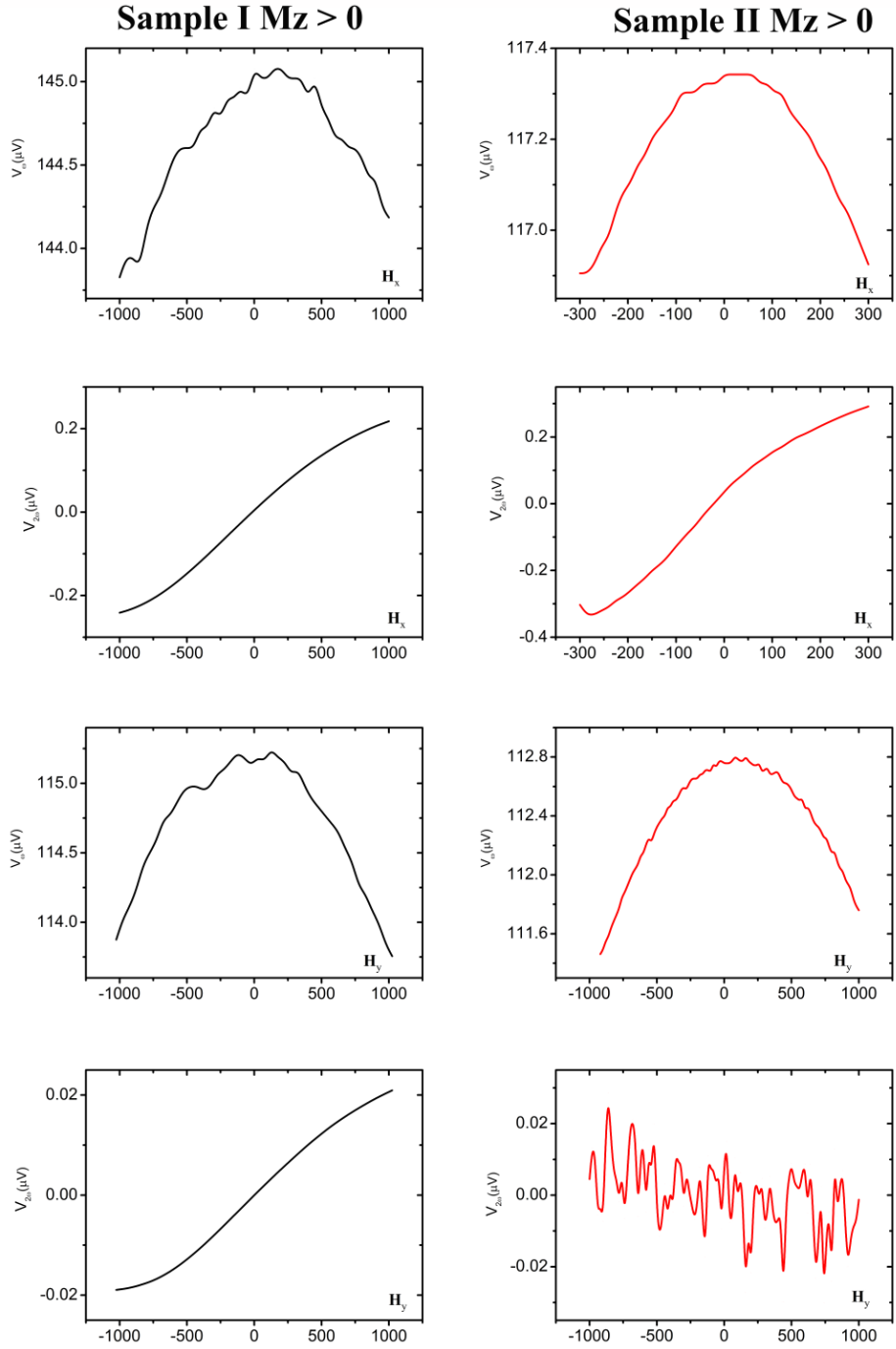

Figure 3 The first ($V_\omega$) and second ($V_{2\omega}$) harmonic Hall voltages of the two samples by sweeping the longitudinal ($H_x$) and transverse ($H_y$) fields. For sample I the frequency is 365 Hz and the amplitude is 2.0 mA, and for sample II the frequency is 315 Hz and the amplitude is 2.0 mA. The results were measured with out-of-plane magnetization component $M_Z>0$.

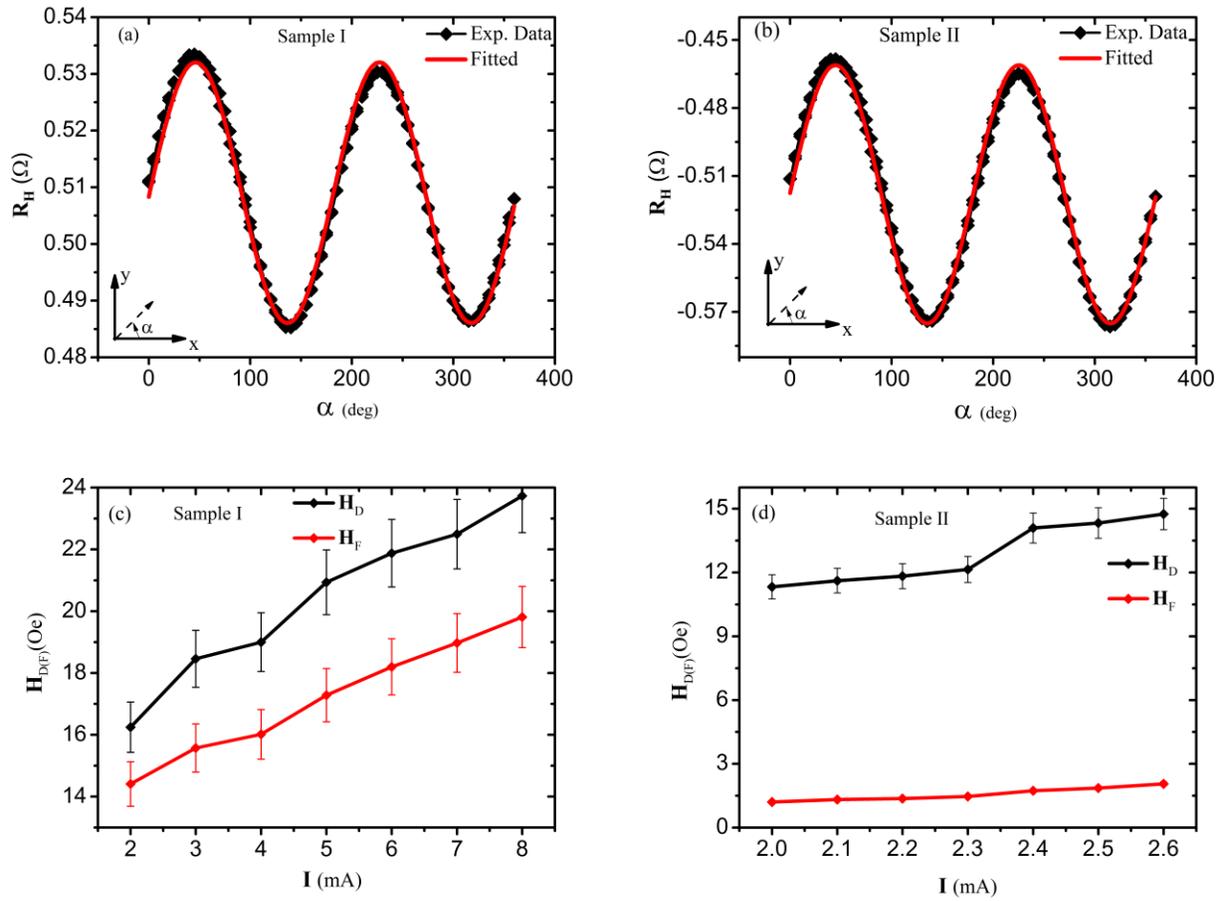

Figure 4 (a) and (b) The $R_H$-$\alpha$ curves of the two samples. Black squares are experimental raw data and the red lines are fitted curves. (c) and (d) The damping like field ($H_D$) and field like effective field ($H_F$) versus applied current of the two samples.